\documentclass[letterpaper, 10 pt, conference]{ieeeconf}  

\IEEEoverridecommandlockouts                              

\usepackage{listings}
\usepackage{amsmath,amssymb}
\usepackage{algorithm}
\usepackage{algorithmic}
\usepackage{hyperref}
\usepackage{booktabs}
\usepackage{tcolorbox}
\tcbuselibrary{breakable}
\usepackage{multirow}
\usepackage{makecell}
\usepackage{tabularx}

\title{\LARGE \bf
Exploration on Demand: From Algorithmic Control to User Empowerment
}

\makeatletter
\providecommand{\bstctlcite}[1]{%
  \@bsphack
  \@for\@citeb:=#1\do{%
    \edef\@citeb{\expandafter\@firstofone\@citeb}%
    \if@filesw
      \immediate\write\@auxout{\string\citation{\@citeb}}%
    \fi
  }%
  \@esphack
}
\makeatother

\author{Edoardo Bianchi$^{1}$
\thanks{*This work was not supported by any organization}
\thanks{$^{1}$Edoardo Bianchi is with Faculty of Engineering,
        Free University of Bozen-Bolzano, 39100 Bozen-Bolzano, Italy
        {\tt\small edbianchi@unibz.it}}%
}

\begin{document}

\bstctlcite{IEEEexample:BSTcontrol}

\maketitle
\thispagestyle{empty}
\pagestyle{empty}

\begin{abstract}

Recommender systems often struggle with over-specialization, which severely limits users' exposure to diverse content and creates filter bubbles that reduce serendipitous discovery. To address this fundamental limitation, this paper introduces an adaptive clustering framework with user-controlled exploration that effectively balances personalization and diversity in movie recommendations. Our approach leverages sentence-transformer embeddings to group items into semantically coherent clusters through an online algorithm with dynamic thresholding, thereby creating a structured representation of the content space. Building upon this clustering foundation, we propose a novel exploration mechanism that empowers users to control recommendation diversity by strategically sampling from less-engaged clusters, thus expanding their content horizons while explicitly exposing the relevance-diversity trade-off. Experiments on the MovieLens dataset demonstrate the system's effectiveness, showing that exploration significantly reduces intra-list similarity from 0.34 to 0.26 while simultaneously increasing unexpectedness to 0.73. Furthermore, our Large Language Model-based A/B testing methodology, conducted with 300 simulated users, reveals that 72.7\% of long-term users prefer exploratory recommendations over purely exploitative ones. Additional relevance metrics, including NDCG@k, Recall@k, and HitRate@k, reveal the expected relevance-diversity trade-off against CF and MMR baselines, positioning the method as a controllable exploration layer for promoting meaningful content discovery.

\end{abstract}

\section{Introduction}

Recommender systems have become integral to digital content platforms, helping users navigate vast catalogs of movies, music, books, and other media. However, traditional recommendation approaches face persistent challenges in balancing personalization with diversity, often leading to over-specialization that limits users' exposure to novel content and creates "filter bubbles" that reinforce existing preferences rather than expanding user horizons.

Collaborative filtering methods, while effective at leveraging collective user behavior, rely heavily on historical interaction data, making them ineffective for new users and prone to popularity bias. Content-based approaches often overemphasize similarity metrics, leading to redundant recommendations that fail to introduce meaningful diversity. Both approaches struggle with the exploration-exploitation trade-off: how to balance familiar, relevant content with novel discoveries that might expand user interests.

The over-specialization problem is particularly pronounced in dynamic content environments where new items are constantly added and user preferences evolve. Traditional recommender systems optimize for immediate user satisfaction, measured through accuracy metrics like precision and recall, which inherently favor recommendations that match past behavior. This optimization strategy, while improving short-term engagement, can lead to recommendation monotony and reduced long-term user satisfaction.

This paper addresses these fundamental limitations through an adaptive recommendation framework that combines semantic clustering with user-controlled exploration. Our approach organizes content using sentence-transformer embeddings and an online clustering algorithm that adapts to content distribution shifts and maintains semantic coherence. The key innovation is an exploration mechanism that allows users to dynamically control the diversity-relevance trade-off in their recommendations, enabling personalized discovery experiences.

To evaluate the system, we report both standard offline relevance and diversity-oriented metrics. We also employ a Large Language Model (LLM)-based user simulation approach that enables scalable A/B testing of recommendation strategies. This methodology addresses a critical limitation of existing evaluation frameworks, which primarily measure how well recommendations match past interactions while failing to capture the value of content discovery and preference expansion.

This work presents the following key contributions:
\begin{itemize}
\item An adaptive online clustering algorithm that groups streaming catalogue items using semantic embeddings and dynamic similarity thresholds;
\item A user-controlled exploration mechanism that increases diversity by sampling from under-explored semantic clusters;
\item An empirical analysis of the relevance-diversity trade-off using NDCG@k, Recall@k, HitRate@k, ILS, Unexpectedness, cold-start, popularity, collaborative filtering, MMR baselines, and an LLM-based A/B study.
\end{itemize}

The remainder of this paper is structured as follows. Section \ref{sec:related} reviews related work on content-based recommendation, clustering approaches, and diversity enhancement techniques. Section \ref{sec:methods} details the proposed system architecture, covering the adaptive clustering algorithm, recommendation strategies, and exploration mechanism. Section \ref{sec:experiments} describes the experimental setup, including dataset preparation, evaluation metrics, and LLM-based user simulation. Section \ref{sec:results} presents comprehensive results and analysis. Section \ref{sec:complexity} discusses computational complexity and scalability considerations. Section \ref{sec:limitations} presents limitations of this work, and Section \ref{sec:conclusions} concludes the paper.

\section{Related Work}
\label{sec:related}

\subsection{Content-Based Recommendation Systems}
Content-based filtering has evolved from metadata-based approaches to sophisticated systems leveraging rich item representations from unstructured content \cite{ContentBasedTrend}. Recent deep learning advances enable more effective content representations, with studies showing that visual features from CNNs \cite{DeepVisFeat} and NLP techniques like word embeddings \cite{nlpRS} significantly enhance semantic content profiling beyond traditional metadata approaches.

\subsection{Clustering-Based Approaches}
Clustering techniques have proven effective for organizing content and enhancing recommendation diversity. Zhang and Hurley \cite{userProfPartitioning} demonstrated that partitioning user profiles into distinct taste segments improves recommendation novelty and balance. Theoretical work by Cohen-Addad et al. \cite{onlineKmeans} provides foundations for online k-means clustering with performance guarantees, while Choromanska and Monteleoni \cite{clustWithExpert} emphasized adaptive mechanisms for streaming data that minimize concept drift while preserving cluster coherence.

\subsection{Diversity and Exploration in Recommendations}
The importance of recommendation diversity is well-established, with Ziegler et al. \cite{ILS} showing that topic diversification improves user satisfaction despite potential accuracy trade-offs. Adamopoulos and Tuzhilin \cite{Unexp} formalized unexpectedness as distinct from novelty and serendipity, providing frameworks for evaluation beyond accuracy metrics. However, most existing exploration approaches lack user control and transparency, which our work addresses through semantic clustering.

\subsection{Evaluation Beyond Accuracy}
Traditional evaluation focuses on accuracy metrics like precision and recall, which are biased toward reinforcing existing preferences \cite{RecSysHand, RecSysHandRicci}. Recent research explores alternative methodologies, including the use of LLMs for simulating human behavior in recommendation scenarios, providing reliable preference judgments that correlate well with human evaluators \cite{simuser, llmrecs}. These approaches build on broader research demonstrating that LLMs can effectively approximate human behavioral patterns and decision-making processes across various domains \cite{llm_trust, agent_hospital}.

\section{Methodology}
\label{sec:methods}

\subsection{System Architecture Overview}
The proposed recommendation system addresses three fundamental challenges in content recommendation: information overload, cold-start scenarios, and limited content diversity. The system employs a modular architecture consisting of five interconnected components designed to work together seamlessly.

The \textit{Item Embedding Module} leverages state-of-the-art sentence transformer models to convert textual metadata into dense semantic representations. The \textit{Adaptive Clustering Module} groups items into semantically coherent clusters that evolve with the content catalog. The \textit{Cold-Start Recommendation Module} enables immediate personalization for new users through keyword-based preference specification. The \textit{Personalized Recommendation Module} generates tailored suggestions for returning users by analyzing interaction history. Finally, the \textit{Exploration Control Module} allows users to dynamically adjust the diversity-relevance balance in their recommendations.

\subsection{Item and User Representation}
\subsubsection{Item Space Definition}

Each item in the system is characterized by comprehensive metadata:
\begin{align}
\text{Item} = \{ID, Title, Tags, Keywords, Description\}
\end{align}

where Tags consist of single descriptive words, Keywords include multi-word phrases, and Description provides a textual summary of content. This rich representation enables nuanced semantic understanding that goes beyond simple keyword matching.

\subsubsection{Semantic Embeddings}
Items are encoded using Sentence-BERT \cite{sentence-bert}, specifically the all-MiniLM-L6-v2 model, which generates 384-dimensional embeddings optimized for semantic similarity tasks. The embedding process concatenates all textual attributes into a unified representation:

\begin{equation}
\mathbf{embedding}_i = \mathrm{SentenceBert}(t_i \oplus g_i \oplus k_i \oplus d_i)
\end{equation}

where $t_i$, $g_i$, $k_i$, and $d_i$ denote the title, tags, keywords, and
description of item $i$, respectively, and $\oplus$ denotes string
concatenation. These embeddings capture contextual relationships and semantic
similarities, enabling meaningful clustering and similarity-based retrieval.

\subsubsection{User Modeling}
Users are represented through their interaction history and preference specifications:
\begin{align}
U &= \{u_1, u_2, \ldots, u_n\} \\
U_i^{prefs} &= \{k_1, k_2, \ldots, k_m\} \\
U_i^{history} &= \{(c_1, v_1), (c_2, v_2), \ldots, (c_n, v_n)\} \\
U_i^{viewed} &= \{id_1, id_2, \ldots, id_p\}
\end{align}

where $U_i^{prefs}$ captures initial keyword preferences, $U_i^{history}$ records interactions as (cluster, item) pairs, and $U_i^{viewed}$ maintains viewed item IDs to prevent re-recommendation.

\subsection{Adaptive Clustering Algorithm}
The adaptive clustering algorithm (Algorithm \ref{algo:clustering}) addresses the challenge of organizing dynamic content catalogs where new items arrive continuously and content distributions may shift over time. Unlike traditional k-means approaches that require fixed cluster numbers, our algorithm dynamically adjusts both cluster membership and similarity thresholds. The algorithm uses cosine similarity for distance computation and maintains cluster centroids as the arithmetic mean of member embeddings. The similarity threshold adapts based on silhouette scores (S) \cite{silhouette}, which provide a measure of clustering quality by comparing intra-cluster tightness with inter-cluster separation:

\begin{align}
distance_{ij} = 1 - \text{cosine\_similarity}(i, j)
\end{align}

Threshold adjustments occur periodically (every 100 items) and follow a conservative strategy: significant reductions when clustering quality is poor ($S < 0.1$), moderate adjustments for suboptimal clustering ($0.1 \leq S < 0.2$), and increases when clusters are well-formed ($S > 0.4$). This approach prevents excessive fragmentation while maintaining semantic coherence.

\begin{algorithm}
\caption{Adaptive Online Clustering}
\label{algo:clustering}
\small
\begin{algorithmic}[1]

\REQUIRE $\mathbf{e}$: item embedding
\REQUIRE id: item identifier
\REQUIRE $\tau$: similarity threshold
\REQUIRE dynamic: dynamic thresholding flag
\REQUIRE $t$: interaction counter
\REQUIRE $f$: update frequency

\STATE \textbf{Cluster Assignment:}
\IF{no clusters exist}
    \STATE create new cluster with $(\mathbf{e}, id)$
\ELSE
    \STATE find nearest centroid $T_n$ using cosine similarity
    \IF{$\mathrm{sim}(\mathbf{e}, T_n) > \tau$}
        \STATE add $id$ to cluster $C_n$
        \STATE update centroid:
        \STATE $T_n \leftarrow \frac{1}{|C_n|} \sum_{i \in C_n} \mathbf{e}_i$
    \ELSE
        \STATE create new cluster with $(\mathbf{e}, id)$
    \ENDIF
\ENDIF

\STATE \textbf{Dynamic Threshold Adjustment:}
\IF{dynamic \AND $(t \bmod f)=0$}
    \STATE compute silhouette score $S$
    \IF{$S < 0.1$}
        \STATE $\tau \leftarrow \max(0.3,\; 0.95\,\tau)$
    \ELSIF{$0.1 \leq S < 0.2$}
        \STATE $\tau \leftarrow 0.98\,\tau$
    \ELSIF{$S > 0.4$}
        \STATE $\tau \leftarrow \min(0.8,\; 1.02\,\tau)$
    \ENDIF
    \STATE merge clusters with similarity $>0.9$
\ENDIF

\end{algorithmic}
\end{algorithm}

\subsection{Cold-Start Recommendation Strategy}
New users face the cold-start problem due to lack of interaction history. Our system addresses this through explicit preference collection during registration, where users select keywords representing their interests.

This approach ensures immediate personalization while collecting initial interaction data. Once sufficient history is available, the system transitions to personalized recommendation mode. The cold-start recommendation strategy is described in Algorithm \ref{algo:coldstart}.

\begin{algorithm}
\caption{Cold-Start Recommendations}
\label{algo:coldstart}
\small
\begin{algorithmic}[1]

\REQUIRE $\mathcal{K}$: user-selected interest keywords
\REQUIRE $k$: number of items to recommend
\REQUIRE $h$: history threshold (min.\ interactions)
\REQUIRE $U_i^{\text{history}}$: user interaction history

\IF{$|U_i^{\text{history}}| \geq h$}
  \STATE \textbf{return} PersonalizedRecommendations$(U_i^{\text{history}}, k)$
\ENDIF

\STATE encode $\mathcal{K}$ into embeddings using SentenceBERT
\STATE compute query embedding: $q \leftarrow \frac{1}{|\mathcal{K}|}\sum_{w\in \mathcal{K}} \mathrm{emb}(w)$
\STATE compute cosine similarity between $q$ and all cluster centroids
\STATE identify top-$m$ clusters with highest similarity scores \hfill ($m{=}3$)
\STATE sample $k$ items uniformly from the identified clusters
\STATE \textbf{return} sampled recommendation list

\end{algorithmic}
\end{algorithm}

\subsection{Personalized Recommendation with Exploration}
For returning users, the system analyzes interaction patterns and provides user-controlled exploration capabilities. The recommendation process balances exploitation of known preferences with exploration of novel content areas.

The exploration mechanism (Algorithm \ref{algo:exploration}) uses an exploration coefficient $\alpha$ to allocate part of the list to clusters where the user has limited engagement, while the remaining positions are sampled from frequently engaged clusters. The original implementation uses $\alpha=2/3$; in the experiments we additionally report a sensitivity analysis over multiple values of $\alpha$.

\begin{algorithm}
\caption{Personalized Recommendations with Exploration}
\label{algo:exploration}
\small
\begin{algorithmic}[1]

\REQUIRE $U_i^{\text{history}}$: user interaction history
\REQUIRE $k$: number of items to recommend
\REQUIRE explore: exploration flag (boolean)
\REQUIRE $\alpha$: exploration coefficient
\REQUIRE $w$: history window size
\REQUIRE $\mathcal{T}$: set of all clusters

\STATE extract recent interactions: $R \leftarrow U_i^{\text{history}}[-w:]$ \hfill
\STATE initialize cluster counts: $M \leftarrow \{\}$

\FOR{each interaction in $R$ with cluster id $c$}
  \STATE $M[c] \leftarrow M[c] + 1$
\ENDFOR

\STATE sort clusters by interaction frequency (descending)
\STATE select top-$m$ clusters from sorted list \hfill ($m{=}3$)
\STATE initialize recommendation list: $\mathcal{R} \leftarrow \emptyset$

\IF{explore}
  \STATE identify remaining clusters: $\mathcal{T}_{\text{other}} \leftarrow \mathcal{T} \setminus \mathcal{T}_{\text{top}}$
  \STATE set exploration budget: $k_{\text{exp}} \leftarrow \lfloor \alpha k \rfloor$
  \STATE sample $k_{\text{exp}}$ items from $\mathcal{T}_{\text{other}}$
  \STATE add sampled items to $\mathcal{R}$
\ENDIF

\STATE set exploitation budget: $k_{\text{explt}} \leftarrow k - |\mathcal{R}|$
\STATE sample $k_{\text{explt}}$ items from top clusters
\STATE add sampled items to $\mathcal{R}$
\STATE \textbf{return} first $k$ items from $\mathcal{R}$

\end{algorithmic}
\end{algorithm}

\section{Experimental Setup}
\label{sec:experiments}
\subsection{Dataset and Preprocessing}
We conducted experiments using the MovieLens 32M dataset \cite{movielens}, which provides extensive user interaction data with rich metadata including movie titles, genres, and user ratings. To ensure computational tractability while maintaining statistical significance, we sampled 20,000 items while preserving the original data distribution across genres and popularity levels.

The preprocessing pipeline concatenated movie metadata (title, genres, and available tags) into unified textual representations suitable for embedding generation. To address representativeness, we also repeat the main $k=10,h=50$ configuration on the full available catalogue after preprocessing, containing 87,585 movies and 32.0M ratings.

\subsection{Implementation Details}
For embedding generation, we employed the Sentence-BERT all-MiniLM-L6-v2 model \cite{sentence-bert}, a 384-dimensional transformer-based encoder fine-tuned on over one billion sentence pairs for semantic similarity tasks. This model provides an effective balance between semantic understanding and computational efficiency. All embeddings were precomputed and cached to minimize runtime overhead.

The clustering configuration initialized the similarity threshold at 0.45, with dynamic adjustments triggered every 100 items. To maintain computational efficiency while preserving accuracy, silhouette scores were computed over random samples of 1000 items, ensuring scalable quality assessment across varying dataset sizes.

Our evaluation framework simulated 300 users with varying interaction histories to assess system performance. Each user received a preference profile constructed by randomly sampling from individual users' rating histories in the MovieLens 32M dataset \cite{movielens}. This sampling preserves natural preference diversity, reflecting realistic user behavior and enabling authentic evaluation across diverse usage scenarios. LLM-based user simulation details are in Section \ref{sec:llm-eval}.

\subsection{Evaluation Methodology}
Because exploration introduces a relevance-diversity trade-off, we evaluate both recommendation accuracy and discovery-oriented properties. Relevance is measured using NDCG@k, Recall@k, and HitRate@k \cite{RecSysHand, RecSysHandRicci,NDCG, RSEval}, with held-out MovieLens ratings $\geq 4.0$ treated as relevant items. Diversity and novelty are measured through ILS \cite{ILS} and Unexpectedness \cite{Unexp}.

\subsection{LLM-Based User Preference Evaluation}
\label{sec:llm-eval}
 To assess user preferences for exploratory versus exploitative recommendations, we conducted simulated A/B testing using the DeepSeek-V3 language model \cite{deepseekv3}. This approach addresses limitations of traditional evaluation methods by incorporating human-like preference judgments that consider factors beyond simple accuracy matching.

For each simulated user, we generated two recommendation sets: one with exploration disabled (exploitation-only) and another with exploration enabled. The LLM was presented with both sets along with the user's interaction history and asked to select the more appealing option. We employed the following evaluation prompt:

\begin{tcolorbox}[
  colback=gray!10,
  colframe=gray!50,
  width=\columnwidth,
  boxsep=2pt,
  left=2pt,
  right=2pt,
  top=2pt,
  bottom=2pt,
  breakable
]
\textbf{System:}\\
You are a movie enthusiast who recently watched: \emph{[user history]}.
You must choose between two recommendation sets.
Consider:
\begin{itemize}
  \item which set better matches your tastes;
  \item which set contains more movies you would actually watch.
\end{itemize}
\textbf{User:}\\
Set~A: \emph{[set A]}.\\
Set~B: \emph{[set B]}.\\
Which set would you choose?
\end{tcolorbox}

Temperature was set to 0.4 to balance response consistency with nuanced variation. This methodology enables large-scale preference assessment that would be prohibitively expensive with human evaluators while providing insights into user behavior patterns.

\subsection{Baseline Comparisons}
We compare against four reference configurations. Cold-start recommendations evaluate the keyword-based initialization strategy. Popularity-based recommendation ranks items by global popularity while incorporating genre preferences derived from user histories. Collaborative filtering uses user-item matrix factorization with 50 latent factors, optimized using alternating least squares. Maximal Marginal Relevance (MMR) \cite{MMR} provides a diversity-aware reranking baseline that explicitly trades off relevance and redundancy. All methods are evaluated using the same users, held-out relevance protocol, and diversity metrics.

\section{Results and Analysis}
\label{sec:results}
Table~\ref{tab:results} presents comprehensive performance results across different experimental configurations. The proposed exploration mechanism obtains the strongest diversity and unexpectedness values, while CF and MMR achieve substantially higher relevance. The results therefore support the method as a controllable exploration layer, not as an accuracy-optimized standalone recommender.

\begin{table*}[t]
    \centering
\caption{Performance comparison across configurations.
Lower ILS indicates higher diversity and higher Unexp indicates increased novelty.
A/B denotes the percentage of users preferring exploratory recommendations.
Relevance metrics are computed from held-out MovieLens ratings $\geq 4.0$.
Results are obtained on 20\,000 items with 300 simulated users; $k$ and $h$ denote list size and history length.}
    \label{tab:results}
    \renewcommand{\arraystretch}{1.25}
    \scriptsize
    \begin{tabular}{cc l c c c c c c}
        \toprule
        \textbf{$k$} & \textbf{$h$} & \textbf{Configuration} &
        \textbf{NDCG} $\uparrow$ & \textbf{Recall} $\uparrow$ &
        \textbf{HitRate} $\uparrow$ & \textbf{ILS} $\downarrow$ &
        \textbf{Unexp} $\uparrow$ & \textbf{A/B(\%)} \\
        \midrule

        \multirow{6}{*}{\makecell[c]{5}} & \multirow{6}{*}{\makecell[c]{10}}
            & Cold Start              & --   & --   & --   & 0.32 & -    & -    \\
        & & Collaborative Filtering & \textbf{0.31} & \textbf{0.09} & \textbf{0.65} & 0.37 & 0.66 & -    \\
        & & Popularity-Based        & 0.20 & 0.05 & 0.48 & 0.47 & 0.61 & -    \\
        & & MMR                     & 0.19 & 0.06 & 0.55 & 0.44 & 0.58 & -    \\
        \cmidrule(lr){3-9}
        & & \textit{Ours} (Exploration Off) & 0.00 & 0.00 & 0.01 & 0.33 & 0.66 & \textbf{51.7} \\
        & & \textit{Ours} (Exploration On)  & 0.00 & 0.00 & 0.01 & \textbf{0.29} & \textbf{0.71} & 48.3 \\

        \midrule

        \multirow{6}{*}{\makecell[c]{10}} & \multirow{6}{*}{\makecell[c]{10}}
            & Cold Start              & --   & --   & --   & 0.37 & -    & -    \\
        & & Collaborative Filtering & \textbf{0.30} & \textbf{0.16} & \textbf{0.80} & 0.38 & 0.66 & -    \\
        & & Popularity-Based        & 0.19 & 0.08 & 0.57 & 0.36 & 0.65 & -    \\
        & & MMR                     & 0.17 & 0.10 & 0.65 & 0.45 & 0.57 & -    \\
        \cmidrule(lr){3-9}
        & & \textit{Ours} (Exploration Off) & 0.00 & 0.00 & 0.03 & 0.34 & 0.66 & \textbf{62.3} \\
        & & \textit{Ours} (Exploration On)  & 0.00 & 0.00 & 0.02 & \textbf{0.28} & \textbf{0.71} & 37.7 \\

        \midrule

        \multirow{6}{*}{\makecell[c]{5}} & \multirow{6}{*}{\makecell[c]{50}}
            & Cold Start              & --   & --   & --   & 0.34 & -    & -    \\
        & & Collaborative Filtering & \textbf{0.39} & \textbf{0.06} & \textbf{0.74} & 0.39 & 0.66 & -    \\
        & & Popularity-Based        & 0.20 & 0.03 & 0.47 & 0.38 & 0.65 & -    \\
        & & MMR                     & 0.23 & 0.03 & 0.65 & 0.43 & 0.61 & -    \\
        \cmidrule(lr){3-9}
        & & \textit{Ours} (Exploration Off) & 0.01 & 0.00 & 0.03 & 0.33 & 0.67 & 27.3 \\
        & & \textit{Ours} (Exploration On)  & 0.00 & 0.00 & 0.01 & \textbf{0.27} & \textbf{0.72} & \textbf{72.7} \\

        \midrule

        \multirow{6}{*}{\makecell[c]{10}} & \multirow{6}{*}{\makecell[c]{50}}
            & Cold Start              & --   & --   & --   & 0.37 & -    & -    \\
        & & Collaborative Filtering & \textbf{0.34} & \textbf{0.10} & \textbf{0.80} & 0.37 & 0.67 & -    \\
        & & Popularity-Based        & 0.18 & 0.05 & 0.63 & 0.36 & 0.65 & -    \\
        & & MMR                     & 0.18 & 0.05 & 0.65 & 0.44 & 0.61 & -    \\
        \cmidrule(lr){3-9}
        & & \textit{Ours} (Exploration Off) & 0.00 & 0.00 & 0.03 & 0.34 & 0.67 & 39.2 \\
        & & \textit{Ours} (Exploration On)  & 0.00 & 0.00 & 0.02 & \textbf{0.26} & \textbf{0.73} & \textbf{60.8} \\
        \bottomrule
    \end{tabular}
\end{table*}

\subsection{Diversity and Novelty Improvements}
The exploration mechanism consistently improves both diversity and novelty across all experimental configurations. Enabling exploration reduces ILS and increases Unexpectedness relative to the exploitation-only version, showing that the system effectively expands recommendation lists beyond the user's most familiar semantic regions. The strongest improvement appears for longer user histories: at $k=10,h=50$, ILS decreases from 0.34 to 0.26, while Unexpectedness increases from 0.67 to 0.73. This suggests that the clustering structure becomes especially useful once sufficient interaction history is available to distinguish habitual clusters from under-explored content areas.

Compared with the baselines, the exploration-enabled configuration achieves the best diversity and novelty results. Collaborative filtering and popularity-based recommendation remain more concentrated around familiar or popular content, while MMR improves diversity through reranking but does not reach the same Unexpectedness values as the proposed exploration mechanism. These results support the main goal of the paper: providing a simple and controllable mechanism for increasing discovery and reducing recommendation redundancy. However, this gain comes with a clear relevance-diversity trade-off, as discussed in the following sections.

\subsection{User Preference Patterns}
The LLM-based A/B testing provides an additional perspective on the effect of exploration. The results suggest that preference for exploration depends strongly on the amount of available user history. For short histories ($h=10$), users tend to prefer the exploitation-only configuration, with 51.7\% preference for exploration-off at $k=5$ and 62.3\% at $k=10$. This is consistent with the intuition that, when little interaction history is available, the system has limited evidence for selecting useful exploratory clusters.

For longer histories ($h=50$), the pattern reverses: exploratory recommendations are preferred in 72.7\% of cases for $k=5$ and 60.8\% for $k=10$. This suggests that exploration becomes more valuable when the system has enough evidence to identify both stable preferences and under-explored semantic regions. Although LLM-based judgments should not be interpreted as definitive human validation, these A/B results indicate that the diversity gains are not merely metric artifacts, but can translate into more appealing recommendation lists..

\subsection{Sensitivity and Larger-Catalogue Validation}
A sensitivity analysis over $\alpha \in \{0.25,0.50,0.66,0.75\}$ confirms that stronger exploration generally lowers ILS and increases Unexpectedness, while relevance remains low. For $k=10,h=50$, ILS decreases from 0.31 at $\alpha=0.25$ to 0.28 at $\alpha=0.75$, while Unexpectedness increases from 0.69 to 0.71. This indicates that $\alpha$ acts as an effective control knob for diversity, but not as a relevance optimizer.

The larger-catalogue experiment confirms the same pattern. On 87,585 movies and 32.0M ratings, at $k=10,h=50$, exploration reduces ILS from 0.33 to 0.28 and increases Unexpectedness from 0.68 to 0.71. CF remains much stronger in relevance, with NDCG@10 of 0.40 and HitRate@10 of 0.82, compared with 0.00 and 0.02 for the exploration-enabled method. Thus, the diversity effect persists at larger scale, but the relevance-diversity trade-off remains substantial.

\subsection{Cold-Start Performance}
The keyword-based cold-start mechanism provides a simple way to initialize recommendations without interaction history. Its ILS values range between 0.32 and 0.37 in the main experiments, indicating moderate diversity and behavior comparable to the non-exploratory personalized setting. This suggests that the same semantic cluster structure can support both initialization and later exploration, although relevance-oriented cold-start evaluation remains future work.

\section{Computational Complexity and Scalability}
\label{sec:complexity}
Cluster assignment requires $O(C \cdot d)$ operations for nearest-centroid search, where $C$ is the number of clusters and $d=384$ is the embedding dimension. Centroid updates scale as $O(|I_c| \cdot d)$ for the modified cluster. Exact silhouette computation is $O(N^2)$ and therefore must be approximated by sampling for large catalogues. Personalized recommendation requires history counting and cluster ranking, approximately $O(h \cdot C + C\log C)$. The larger-catalogue run with 87,585 movies confirms that the implementation can be evaluated beyond the 20,000-item subset, although the current method still requires stronger item-level ranking to improve relevance.

\section{Limitations and Future Work}
\label{sec:limitations}
The main limitation is that exploratory items are selected primarily at the cluster level, with weak item-level relevance ranking. This explains the low NDCG, Recall, and HitRate compared with CF and MMR. Future versions should combine semantic cluster-level exploration with a stronger relevance model, for example by ranking candidates inside each selected cluster using collaborative-filtering scores or learned user-item relevance. A second limitation is that the LLM-based A/B test is not validated against real users; it should therefore be interpreted as a scalable preference-oriented evaluation rather than definitive evidence of human satisfaction. Finally, the experiments focus on MovieLens; cross-domain validation remains necessary.

\section{Discussion and Conclusion}
\label{sec:conclusions}
This paper presented a semantic clustering framework for user-controlled exploration in recommender systems. The final evaluation shows a clear and important trade-off. On the positive side, the proposed exploration mechanism consistently reduces intra-list similarity and increases unexpectedness across list sizes, history lengths, and a larger-catalogue setting. On the negative side, standard relevance metrics show that the current implementation is not competitive with CF or MMR as a standalone recommender.

The main value of the approach is therefore not accuracy replacement, but controllable discovery. By exposing an exploration coefficient over semantic clusters, the method provides an interpretable mechanism for increasing diversity and reducing recommendation redundancy. The LLM-based A/B study further suggests that this form of exploration can be attractive for longer-history users, even though human validation remains necessary. These results indicate that cluster-level exploration can be useful as an add-on layer to relevance-oriented recommenders, especially when user agency and exposure diversity are design goals. Future work should integrate stronger item-level ranking and validate exploratory preferences with human users.


\bibliographystyle{IEEEtran}
\bibliography{sample-ceur}

\end{document}